\documentclass[final,3p,times,twocolumn]{elsarticle}

\usepackage{amssymb}
\usepackage{amsmath}
\usepackage{graphicx}
\usepackage{caption}
\usepackage{subcaption}
\usepackage[colorlinks=true, urlcolor=blue, citecolor=blue, linkcolor=blue]{hyperref}

\journal{Microelectronic Engineering}

\begin{document}

\begin{frontmatter}

\title{\vskip -40pt \sf{\small This is an author-created preprint of an article published in Microelectronic Engineering {\bf 215} (2019) 110982 (6pp).\\ The Version of Record is available online at \href{https://doi.org/10.1016/j.mee.2019.110982
}{https://doi.org/10.1016/j.mee.2019.110982}.}
\vskip 10 pt \hrule \vskip 40pt
Nano-patterning of cuprate superconductors by masked He$^+$ ion irradiation: 3-dimensional profiles of the local critical temperature}

\author{K. L. Mletschnig}
\author{W. Lang\corref{cor1}}
\cortext[cor1]{Corresponding author: W. Lang}
\ead{wolfgang.lang@univie.ac.at}
\address{University of Vienna, Faculty of Physics, Electronic Properties of Materials, Boltzmanngasse 5, A-1090, Wien, Austria}

\vspace{10pt}

\begin{abstract}
Irradiation of cuprate high-$T_c$ superconductors with light ions of moderate energy creates point defects that lead to a reduction or full suppression of the critical temperature. By shaping the ion flux with a stencil mask, nanostructures for emerging superconducting electronics can be fabricated. The 3-dimensional shape of such defect landscapes is examined, based on calculations of full collision cascades and atom displacements. A relation between the calculated defect density and experimental values of the critical temperature $T_c$ in thin YBa$_2$Cu$_3$O$_{7-\delta}$ films is etablished that allows to determine the distribution of local $T_c$'s and its 3-dimensional visualization. The results confirm that, using 75~keV He$^+$ ion irradiation and a stencil mask, well-defined patterns of non-superconducting material in the superconducting matrix can be produced with low blurring.

\end{abstract}


\end{frontmatter}

\section{Introduction}

The cuprate high-$T_c$ superconductors (HTSCs) are an important class of materials for both fundamental science research, as well as novel technological applications. Obviously, a primary reason is the rather high critical temperature $T_c$, which allows to operate such devices with reliable and easy-to-handle cryocooler technology. The well-known applications of almost loss-less energy transport in superconducting cables and the creation of intense magnetic fields by superconducting coils require high critical current densities \cite{CARD03M} that can be achieved for instance by inclusion of nanoparticles \cite{HAUG04} to increase flux pinning. On the other hand, possible applications of HTSCs in superconducting circuits and devices for novel quantum technologies \cite{WORD17M,MOSH10M} do not require high critical currents but instead advanced nanostructuring methods to pattern thin films of HTSCs. However, the anisotropic layered structure and the brittle nature of HTSCs pose severe constraints to conventional micro- and nanopatterning methods.

But it is exactly the complex crystal structure of HTSCs that allow for an intriguing and unconventional nanopatterning method. In the prototypical HTSC YBa$_2$Cu$_3$O$_{7-\delta}$ (YBCO), oxygen is situated at several different atomic positions. Whereas the oxygen atoms in the CuO$_2$ planes that carry the supercurrent have a binding energy of about 8.4 eV \cite{TOLP96}, the oxygen atoms in the CuO chains are bound by only about 1 eV \cite{CUI92}. They are also important for superconductivity, because they supply the charge transfer to the CuO$_2$ planes. For this reason, YBCO is rather sensitive to light-ion irradiation during which the oxygen --- and to a lesser extent also the other atoms --- are displaced from their sites. The resulting vacancies and interstitials are point defects in an otherwise still intact crystallographic framework. Due to the $d$-wave symmetry of the superconducting gap in the HTSCs, even minor distortions of the atomic lattice lead to a reduction  of $T_c$ \cite{WANG95b}, in particular when the defects concern the CuO$_2$ planes \cite{TOLP96}. The delicate influence of the oxygen positions in the YBCO structure is for instance illustrated by the observations that $T_c$ can be increased again when disordered oxygen is re-ordered by thermal annealing \cite{JORG90} or by optical illumination \cite{STOC98a}. Thus, irradiation with light ions is a convenient tool for a controlled modification of the  superconducting properties of YBCO \cite{LANG04,CYBA14a}.

\begin{figure*}[t!]
\centering
    \begin{subfigure}[b]{0.52\textwidth}
       \centering
       \includegraphics[width=\textwidth]{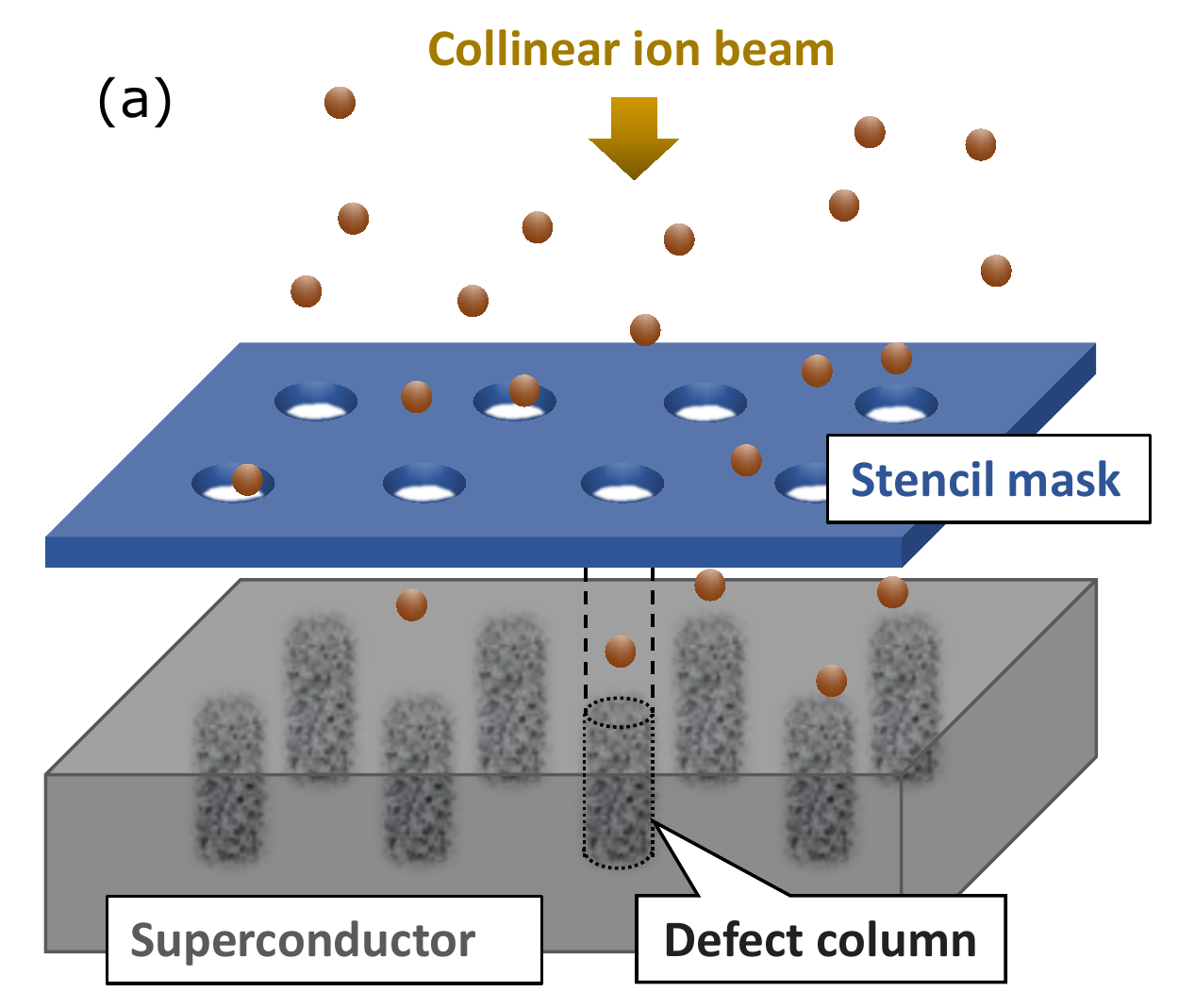}
    \end{subfigure}
    \hfill
    \begin{subfigure}[b]{0.45\textwidth}
       \centering
    \includegraphics[width=\textwidth]{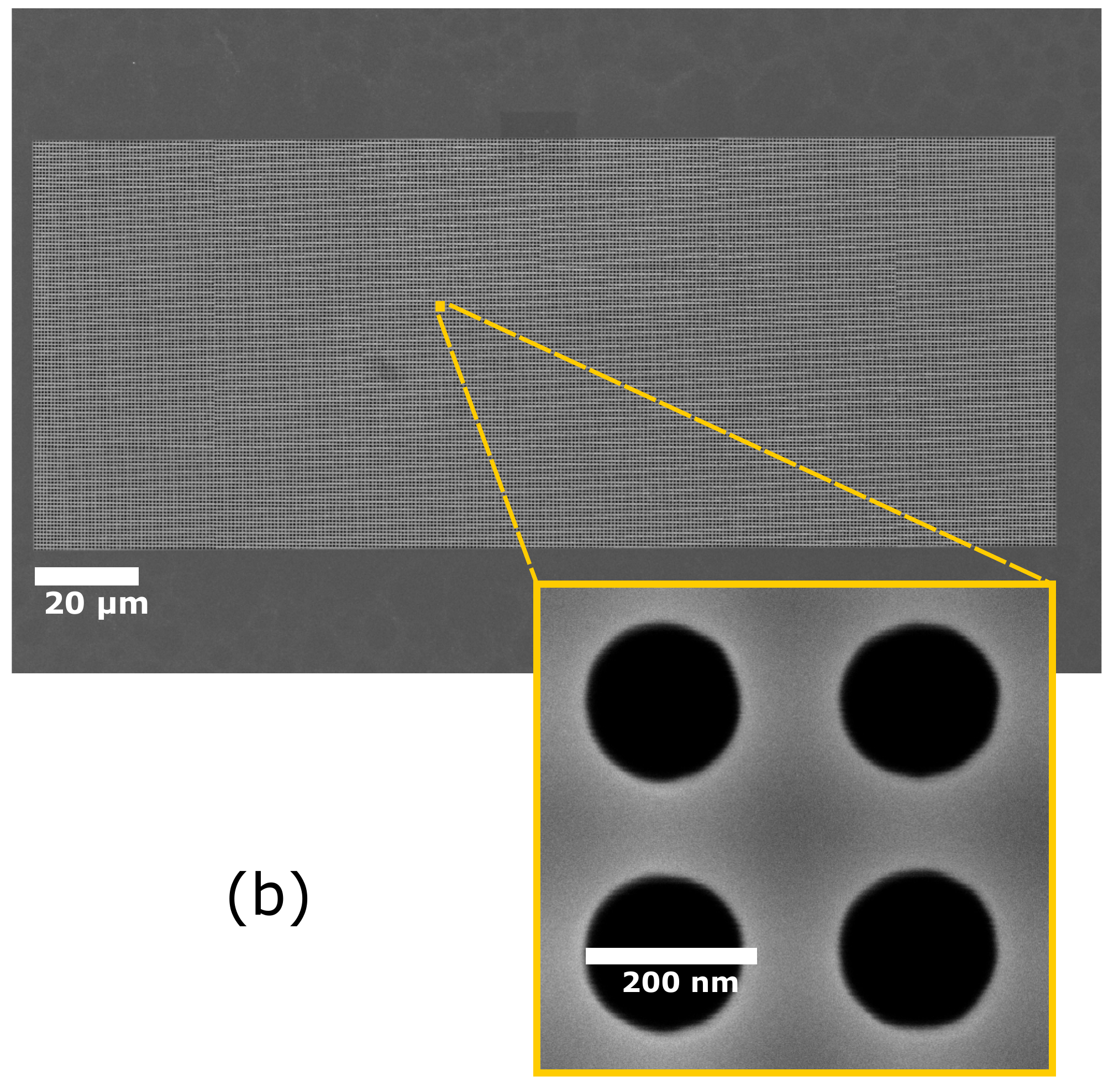}
    \end{subfigure}
\caption{(a) The principle of masked ion beam direct structuring (MIBS), where a collinear ion beam is partially blocked by a stencil mask. Ions reaching the superconductor's surface create columns of point defects that reduce or suppress the critical temperature. (b) Scanning electron microscopy picture of a typical 2 $\mu$m-thick silicon stencil mask used for MIBS. It consists of a rectangular area of $200\,\mu\text{m} \times 80\,\mu\text{m}$ perforated with a square array of holes with $(302 \pm 2)\,$nm pitch. The blow-up shows a detailed view of a few holes with diameters $(180 \pm 5)\,$nm \cite{HAAG14}.}
\label{fig:MIBS}
\end{figure*}

Nanopatterning using stencil-mask lithography has been applied to several materials and has a number of advantages over other structuring techniques \cite{VAZQ15,VILL11a}. In combination with the afore-mentioned specific properties of HTSCs, shadowed ion irradiation can be employed to locally modify the superconducting properties. This provides a route to create small non-superconducting structures in thin YBCO films that can act as commensurate pinning centers for magnetic flux quanta.

Some of the previous experiments have used a variant of the stencil technique by using a pre-patterned layer of photoresist with etched voids \cite{KAHL98,KATZ00,BERG05,SWIE12} or a deposited metal layer that is patterned by ion beam milling \cite{KANG02a,BLAM03} as the ion-blocking material. By irradiating YBCO with 200 keV Ne$^+$ \cite{KATZ00} or 100 keV O$^+$ \cite{BERG05} ions through a 20~nm wide slit in the photoresist the fabrication of Josephson junctions has been demonstrated. However, such procedures lack some of the typical advantages of stencil lithography.

In our approach, a thin Si stencil mask is placed on top of the YBCO film and kept at well-defined distance by a  $1.5\text{-}\mu$m-thick spacer layer. By this procedure any contact between the surfaces of the mask and the YBCO film is avoided. This masked ion beam direct structuring method (MIBS), sketched in Fig.~\ref{fig:MIBS}(a), provides a 1:1 projection of the hole pattern inscribed in the mask, since the ion beam from a commercial implanter is highly collinear and can only reach the YBCO surface through the mask holes. Scanning electron microscopy pictures of a typical mask are presented in Fig.~\ref{fig:MIBS}(b). The ion irradiation produces columnar defect-rich regions (CDs) in the sample. A detailed description of MIBS can be found elsewhere \cite{LANG06a,ZECH17a}. Patterns with linewidths down to 67~nm have been demonstrated \cite{PEDA10}.

Ion-irradiation has many advantages for nanopatterning of HTSCs. By adjusting the ion fluence, the $T_c$ of the superconductor can be reduced, it can be converted to a normal conductor, or even to an insulator. With the ion's energy the penetration depth can be tuned according to the thickness of the HTSC film. In contrast to chemical etching or ion milling, the surface of the HTSC remains essentially flat after ion irradiation, which avoids deterioration of the superconducting part of the film by out-diffusion of oxygen through the open side faces. It is a severe stability issue when many small structural features would be prepared by removing material.

Using ion irradiation together with stencil lithography avoids any contact with the sample surface, does not require chemical treatment or etching, is a time-economic parallel method applicable to large areas, and the mask can be reused many times. Another non-contact method has been recently demonstrated employing a focused He$^+$ beam. Josephson junctions in thin YBCO bridges are fabricated by irradiating narrow stripes with the focused 30~keV ion beam of a helium ion microscope that is scanned over the sample surface \cite{CYBA15,CHO18,CHO18a,MULL19}. Since no mask is used this method offers great flexibility and sub-10~nm resolution \cite{MULL19}, but can be only applied to very thin HTSC films due to penetration depth limitations of the available ion energy.

\begin{figure}[t!]
 \centering
 \includegraphics[width=\columnwidth]{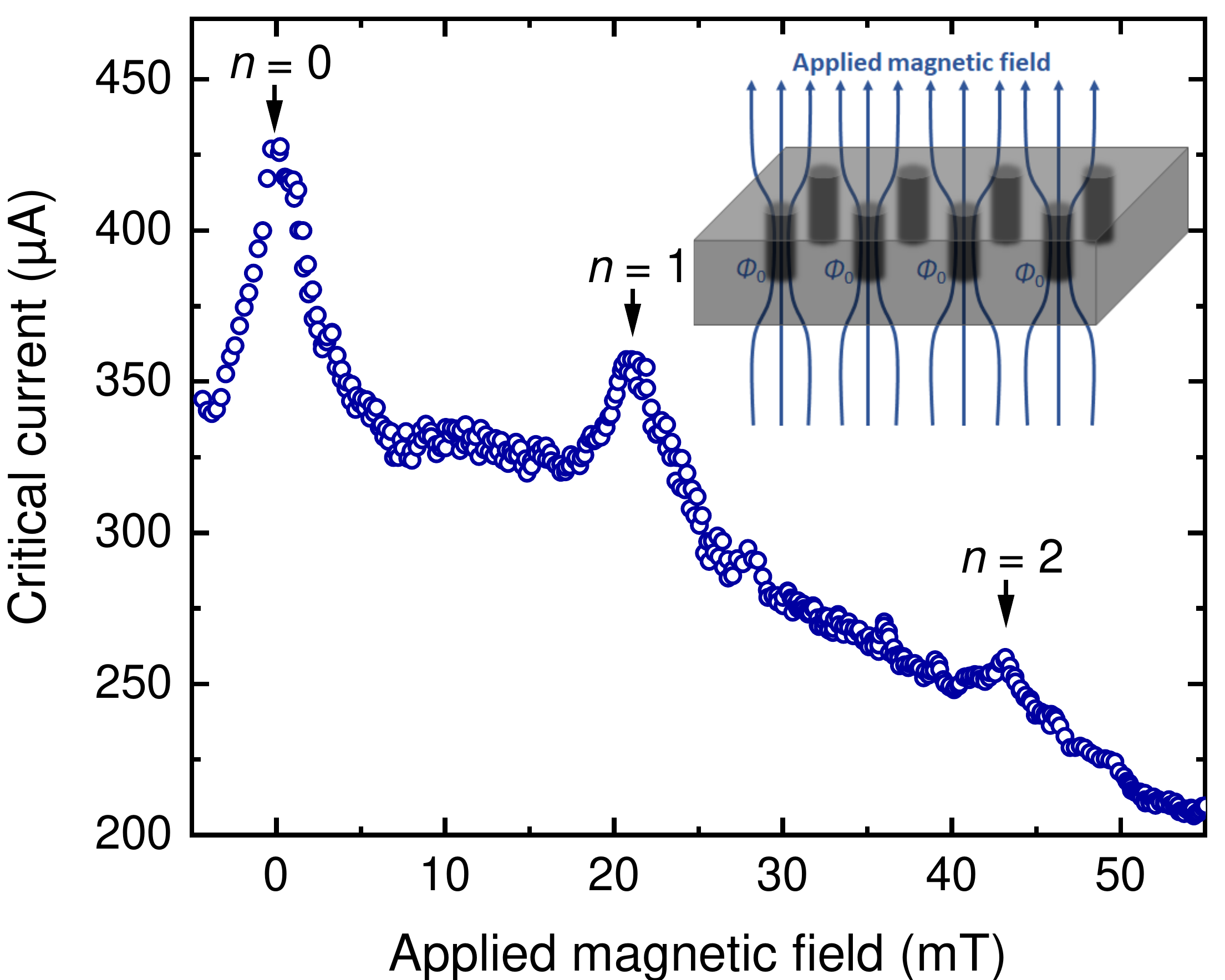}
 \caption{Critical current of a YBCO sample, nanopatterned with a square array of defect columns by MIBS, as a function of the applied magnetic field $B$. Each individual datum was measured after in-field cooling from above $T_c$ to 35~K to achieve the ground state of the vortex arrangement. The inset shows a sketch of the magnetic field profile at the matching field $B_m = 22.6$~mT, corresponding to the $n=1$ peak of the critical current, where every defect column is occupied by one flux quantum $\phi_0$. The current is aligned parallel to the defect rows. Data taken from \cite{ZECH17a}.}
\label{fig:Matching}
\end{figure}

Using MIBS with a 75 keV He$^+$ ion beam with a fluence of $3 \times 10^{15}\ \textrm{cm}^{-2}$, supplied by a commercial ion implanter, a square array of CDs with lattice constant $d = (302 \pm 2)$~nm was patterned into a $(210 \pm 10)$~nm thick YBCO film on a MgO substrate. The thin Si stencil mask (custom fabricated with e-beam lithography by ims-chips, Germany) was perforated with a square array of about $670 \times 270$\ holes with diameters $D = (180 \pm 5)$~nm and covered the entire sample area used for the electrical measurements. Marker holes in the stencil mask were used to align the hole array in an optical microscope parallel to the YBCO bridge with a deviation of less than $0.3^\circ$. Details of the fabrication method have been reported elsewhere \cite{ZECH17a}.

The desired local suppression of $T_c$ in thin YBCO films can be demonstrated by commensurability effects of magnetic flux quanta $\phi_0 = h/(2e)$, where $h$ is the Planck constant and $e$ the electron charge, the so-called fluxons or vortices \cite{HAAG14}. At the matching field $B_m = \phi _0/d^2$ exactly one fluxon is trapped in every CD which leads to a peak in the critical current as shown in Fig.~\ref{fig:Matching}. Besides this $n=1$ peak, it is possible to observe peaks when a larger number $n$ of fluxons are trapped in every CD, or when no magnetic field is present ($n=0$). Since the matching field $B_m$ is solely determined by the density of CDs, the peak in the critical current at $B_m$ provides direct evidence of the commensurate trapping of fluxons in the irradiated areas. Details on the observation of fluxon commensurability effects in the critical current of YBCO films, patterned with  artificial defects by MIBS, are reported elsewhere \cite{HAAG14}. In addition, an unconventional critical state \cite{ZECH17a} and a sign change of the Hall effect \cite{ZECH18a} have been observed in such samples.

Although it is well established that fluxons can be trapped by regions in which superconductivity is suppressed, little is known about the 3-dimensional (3D) profile of the local critical temperature $T_c$ in and around CDs fabricated by ion irradiation. The evaluation of the shape of the irradiated columns and the possible blurring with increasing depth is important for an in-depth understanding of the experimental results. In this paper, we present simulations of defect cascades after ion irradiation through a stencil mask and an assessment of the resulting local $T_c$ variation by comparing the calculated defect density to experimental data.

\section{Methods}

The defects caused by the impact of ions onto the surface of the superconductor are calculated with the program ``Transport of Ions in Matter'' (TRIM) \cite{ZIEG85M}, which is part of the package ``The Stopping and Range of Ions in Matter'' (SRIM) \cite{ZIEG10}. It can compute the impact of ions on solids using a binary collision approximation of ion-atom and atom-atom collisions, providing the energy loss and final spatial distribution of ions, and, most importantly, the full collision cascades. It is based on approximations to quantum mechanical screened interaction potentials and a statistical Monte Carlo algorithm to compute nuclear and electronic stopping cross-sections in amorphous targets. However, ion channeling, thermal effects, diffusion and recrystallization are not considered.

Our simulations were performed with the \emph{Monolayer Collision Steps} method to force a calculation of ion-atom interaction and full interatomic collision cascades in each target layer, thus disabling the free-flight-path approximation. To enable comparison with our previous experimental results, the incident ion species is He$^+$ with an energy of 75~keV and a direction orthogonal to the sample surface. The density of the YBCO target layer is set to 6.3~g~cm$^{-3}$ \cite{MARW89a} with the number of compound atoms according to their  stoichiometric abundance, but considering the one oxygen site per elementary cell in the CuO chains separately. The displacement energies $E_d$ for the various atom species are: $E_d^{O_c}=1$~eV for the chain oxygen atoms \cite{CUI92}, $E_d^{O_p}=8.4$~eV for the in-plane and apical oxygen atoms \cite{TOLP96}, $E_d^{Cu}=15$~eV for the Cu atoms \cite{LEGR93}, and 35~eV for the Y and Ba atoms. Note that the results are not sensitive to minor deviations from these values. For further evaluations, displacements of all atoms in the YBCO structure are considered as defects, where O atoms represent about 75~\% and Cu atoms about 18~\% of the total defect number.

The output of SRIM has to be further processed to reflect the actual experimental situation. In SRIM, collision cascades are calculated independently for individual ions impacting at the same spot at coordinates $(x,y)=(0,0)$. The number of ions $N$ that correspond to a particular irradiation fluence $\Phi$ of a circular area with diameter $d$ is $N=\Phi {d^2}\pi/4$. For typical experimental values $d=180$~nm and $\Phi=3\times 10^{15}\textrm{~cm}^{-2}$, and hence $N=763,407$. For each incident ion, the impact coordinates $(x_i,y_i)$ are determined by evenly distributed random number pairs with $\sqrt{x_i^2+y_i^2}\leq d/2$ and all output data from SRIM (defects) related to this ion trajectory are  then shifted by $(x_i,y_i)$.

\begin{figure}[t!]
    \centering
   	\includegraphics[width=\columnwidth]{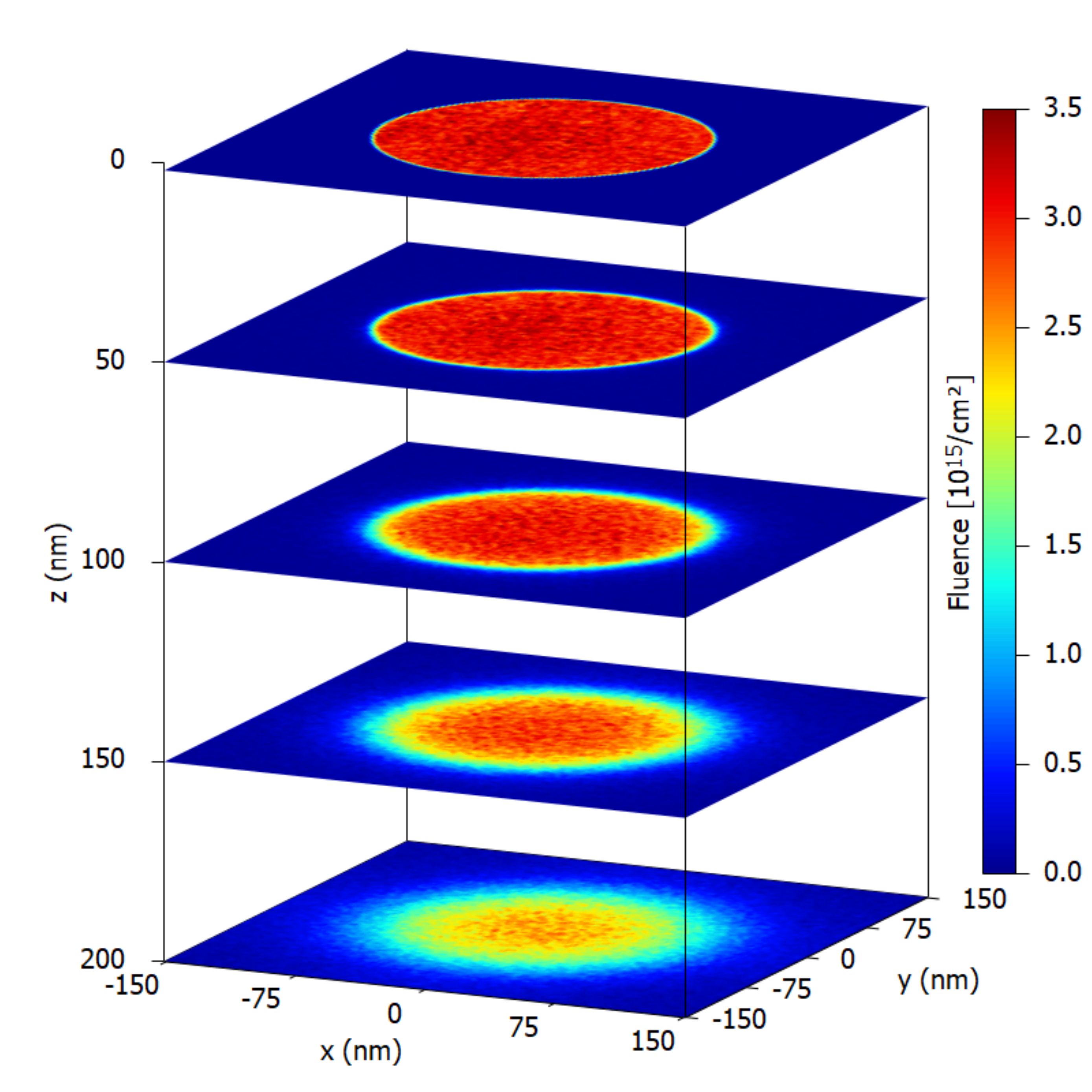}
    \caption{Ion fluence profiles at various depths for a YBCO thin film uniformly irradiated with 75~keV He$^+$ ions through a stencil mask hole with 180 nm diameter.}
    \label{fig:fluence}
\end{figure}

Since SRIM uses a Monte-Carlo method on an amorphous target structure, details of the elementary cell are not resolved and, thus, the various defects are counted in cells of $2 \times 2 \times 2$~nm$^3$ \--- a size that roughly corresponds to the in-plane Ginzburg-Landau coherence length of YBCO. The results are then converted to commonly used \emph{defects per atom} (dpa) values considering the YBCO's crystal structure. In our samples, the CDs form a square array of CDs (CDA) with 300~nm lattice constant. If an ion trajectory leaves the CDA unit cell and, thus, produces defects in the neighboring CD, it is mapped back to the primary unit cell for proper visualization.

\section{Results and discussion}

\begin{figure*}[t!]
    \centering
    \begin{subfigure}[b]{0.48\textwidth}
        \centering
        \includegraphics[width=\textwidth]{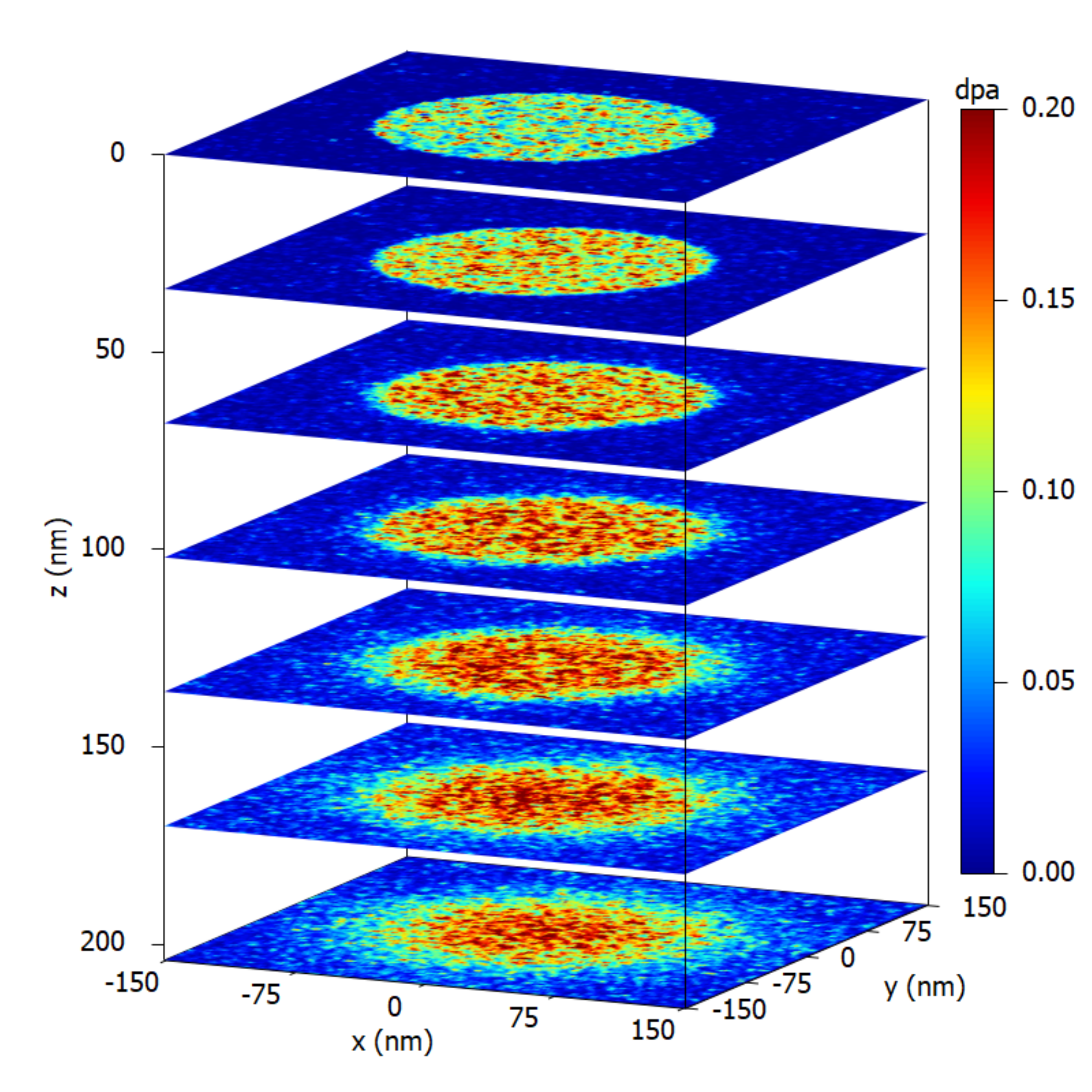}
		\subcaption{Lateral profiles at various depths.}
	\end{subfigure}
    \hfill
    \begin{subfigure}[b]{0.48\textwidth}
        \centering
        \includegraphics[width=\textwidth]{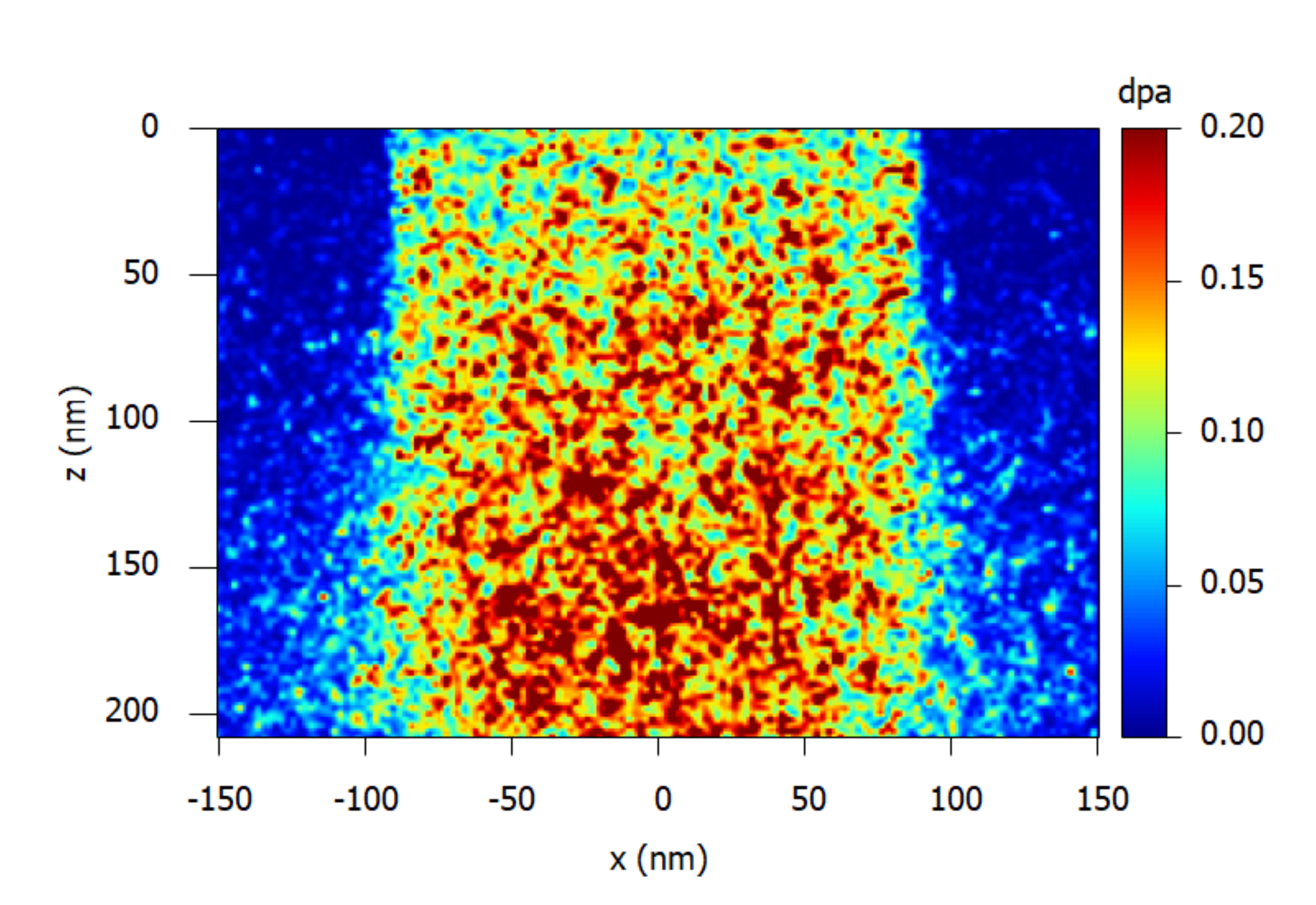}
        \subcaption{Cross-sectional view.}
    \end{subfigure}
    \caption{Distribution of the defect density (in dpa) within and around a defect column produced by 75~keV He$^+$ ion irradiation of YBCO with a fluence of $3\times10^{15}\textrm{~cm}^{-2}$.}
    \label{fig:dpa}
\end{figure*}

The fluence of He$^+$ ions at various depths of a 200-nm-thick YBCO film is presented in Fig.~\ref{fig:fluence}. With increasing depth, the ions are multiply deflected from their previous trajectories, which leads to a blurring effect and to a change from the homogeneous fluence at the top of the sample to a peaked distribution towards the bottom. At first sight, this straggling of ion trajectories and the resulting reduction of their fluence seems to impede the envisaged applications.

However, for the calculation of the local dpa values not only the ion trajectories but also the full collision cascades resulting from the knock-ons of target atoms have to be considered. The resulting dpa profile is shown in Fig.~\ref{fig:dpa} for various depths and in a cross-sectional view. Surprisingly, the defect density is lowest near the sample surface that is subjected to the ion beam and gradually increases with depth as reported before \cite{CYBA14a}. This is a direct consequence of the increasing number of additional collision cascades originating from knocked-on target atoms. The dpa profile near the bottom of the sample is also somewhat blurred, but as can be seen in Fig.~\ref{fig:dpa}(b) the CD is well formed over the entire depth and its diameter increases by about 20~nm.

Several mechanisms have to be considered for understanding the depth dependence of the dpa values. On the one hand, the dpa increase due to accumulation of recoil cascades. Assuming a constant ion fluence and negligible backscattering of ions and impinged target atoms, the collision cascades originating from primary recoil atoms create a plume of defects around their trajectory, but also trigger secondary collision cascades by hitting other atoms, and so on. Thus, the density of collision cascades increases with depth, in particular at the center of the irradiated area. Additionally, the ions lose energy with depth due to electronic stopping, i.e., target ionization. With reduced energy the nuclear stopping cross-section of ions increases and so does the number of created defects. All these processes cause an increase of dpa with depth.

\begin{figure}
\centering
\includegraphics[width=\columnwidth]{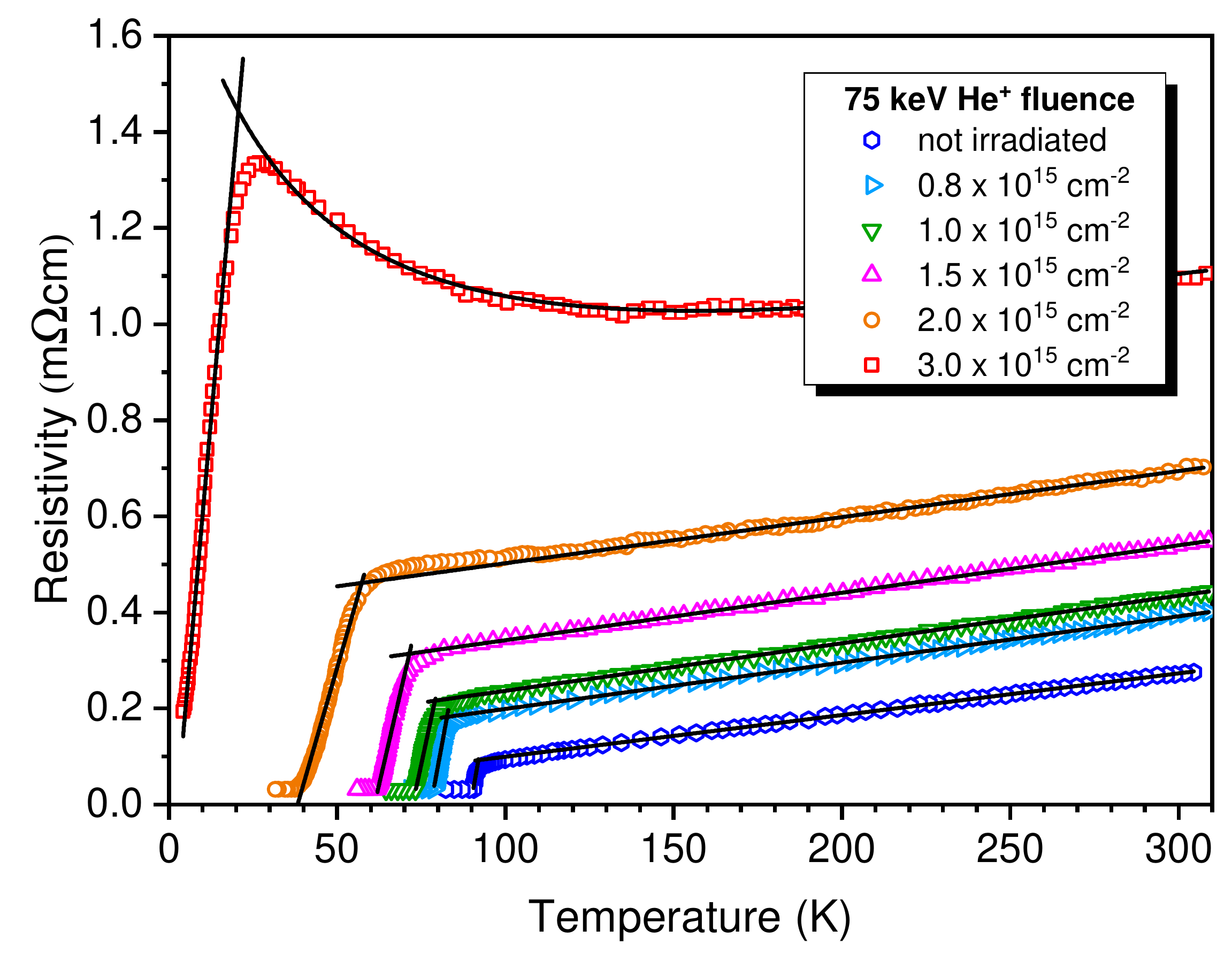}
\caption{Temperature dependence of the resistivity and superconducting transition of a full-area irradiated YBCO film for several 75~keV He$^+$ ion fluences. At a fluence of $4\times 10^{15}$~cm$^{-2}$, superconductivity is fully suppressed (not shown). The lines and their intersection indicate the evaluation of $T_\text{c-onset}$. Data taken from \cite{LANG04}.}
\label{fig:Tc}
\end{figure}

On the other hand, the ion fluence decreases with depth, as can be seen in Fig.~\ref{fig:fluence}, which counteracts the increase of dpa. In samples with a larger thickness than those studied here, this would eventually lead to the implantation of ions, the reduction of dpa, and finally, limiting the defect creation process at some depth. For most applications this is not acceptable and, hence, the ion energy must be high enough to ensure (almost) full penetration of the sample. By choosing ion energy and fluence the defect profile can be adjusted with the constraints that higher energy requires higher fluence to achieve the same dpa, but in turn provides larger penetration depth.

The main aim of this paper is to estimate the 3-dimensional $T_c$ profile within and around the CDs. To this end the defect density from the SRIM simulation has to be correlated with experimental $T_c$ data. The issue that the simulated dpa values are expected to be higher than those in the real system, since several processes for defect annihilation are not considered in SRIM, can be overcome by mapping the simulation data to experimental $T_\text{c-onset}$ results from our previous work on full-area irradiation of thin YBCO films with 75~keV He$^+$ ions \cite{LANG04} shown in Fig.~\ref{fig:Tc}. $T_\text{c-onset}$ marks the temperature above which the entire sample is in the normal state.

The reduction of $T_c$ with increasing point defect density can be modeled by the pair-breaking theory of Abrikosov and Gor'kov \cite{ABRI60}. Although devised for magnetic impurities, the theory can be also used to describe the situation in HTSCs, where non-magnetic point defects can break up superconducting pairs due to the $d$-wave symmetry of the superconducting gap \cite{LESU90,TOLP96,TOLP96a}. The normalized $T_\text{c-onset}/T_0$, where $T_0=90$~K is the critical temperature of the non-irradiated superconducting film, can be expressed as a function of the pair-breaking scattering time $\tau_p$ \cite{TOLP96}
\begin{equation}
\label{eq:pairbreaking}
\ln\Big(\frac{T_{c\text{-onset}}}{T_0}\Big)=\Psi\Big(\frac{1}{2}\Big) - \Psi\Big(\frac{1}{2}+\frac{C}{4 \pi \tau_p T_{c\text{-onset}}} \Big),
\end{equation}
where $\Psi$ is the digamma function, $C$ a constant of dimension K$\cdot$s and $\tau_p$ is inversely  proportional to the defect density expressed in dpa.

\begin{figure}[t]
\centering
\includegraphics[width=\columnwidth]{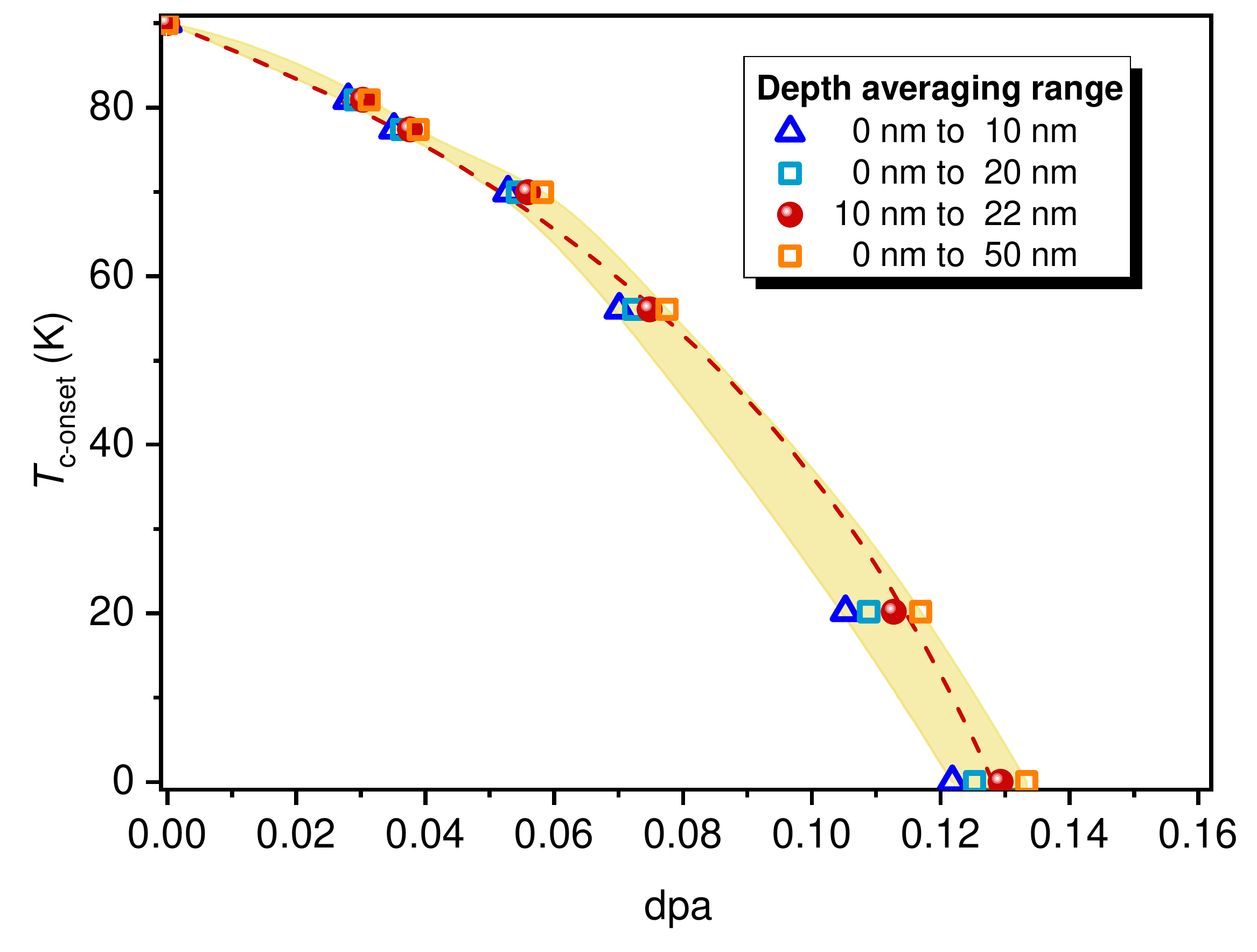}
\caption{Relation between experimentally determined $T_{c-\text{onset}}$ values and simulated defect density in dpa averaged over different cylindrical slices. The symbols indicate different depth zones for computing the mean dpa. The dashed curve is the fit to the data for 10~nm to 22~nm depth. The yellow region marks a possible range for the fit, explained in the text.}
\label{fig:dpa_Tc}
\end{figure}

In Fig.~\ref{fig:dpa_Tc} the relation between the experimentally determined $T_\text{c-onset}$ values and the defect density obtained from the simulations is presented. $T_\text{c-onset}$ is determined by the layer in the sample with the lowest defect density and, thus, the highest local $T_c$. Since the dpa vary with the depth in the sample, as illustrated in Fig.~\ref{fig:dpa}(b), it makes a difference where the dpa value is determined. The previous findings indicate that the dpa is lowest near the surface of the sample, but in the experiments it has also to be considered that a top layer of a few nm thickness is deteriorated by ambient influences. It is therefore reasonable to average the dpa in a slice from 10~nm to 22~nm depth, which corresponds to a thickness of about 10~unit cells of YBCO. As can be noted in Fig.~\ref{fig:dpa_Tc} this choice is not critical, as long as the upper part of the sample is used. The qualitative variation of $T_\text{c-onset}$ with dpa is similar to results obtained with 190~keV He$^+$ irradiation \cite{NEDE89,LESU90}.

\begin{figure*}[t!]
    \centering
    \begin{subfigure}[b]{0.48\textwidth}
        \centering
        \includegraphics[width=\textwidth]{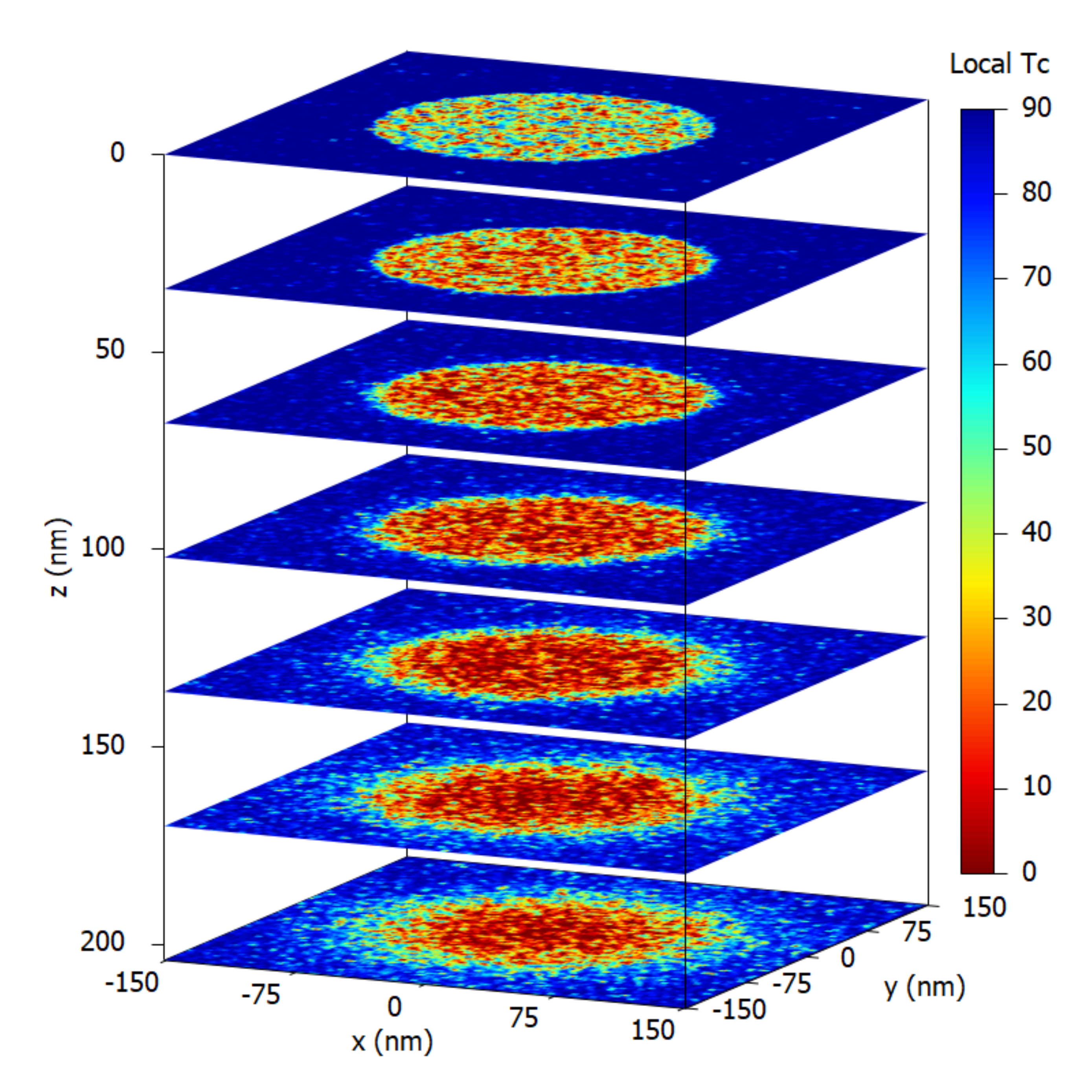}
        \subcaption{Lateral local $T_c$ profiles at various depths.}
	\end{subfigure}
    \hfill
    \begin{subfigure}[b]{0.48\textwidth}
        \centering
        \includegraphics[width=\textwidth]{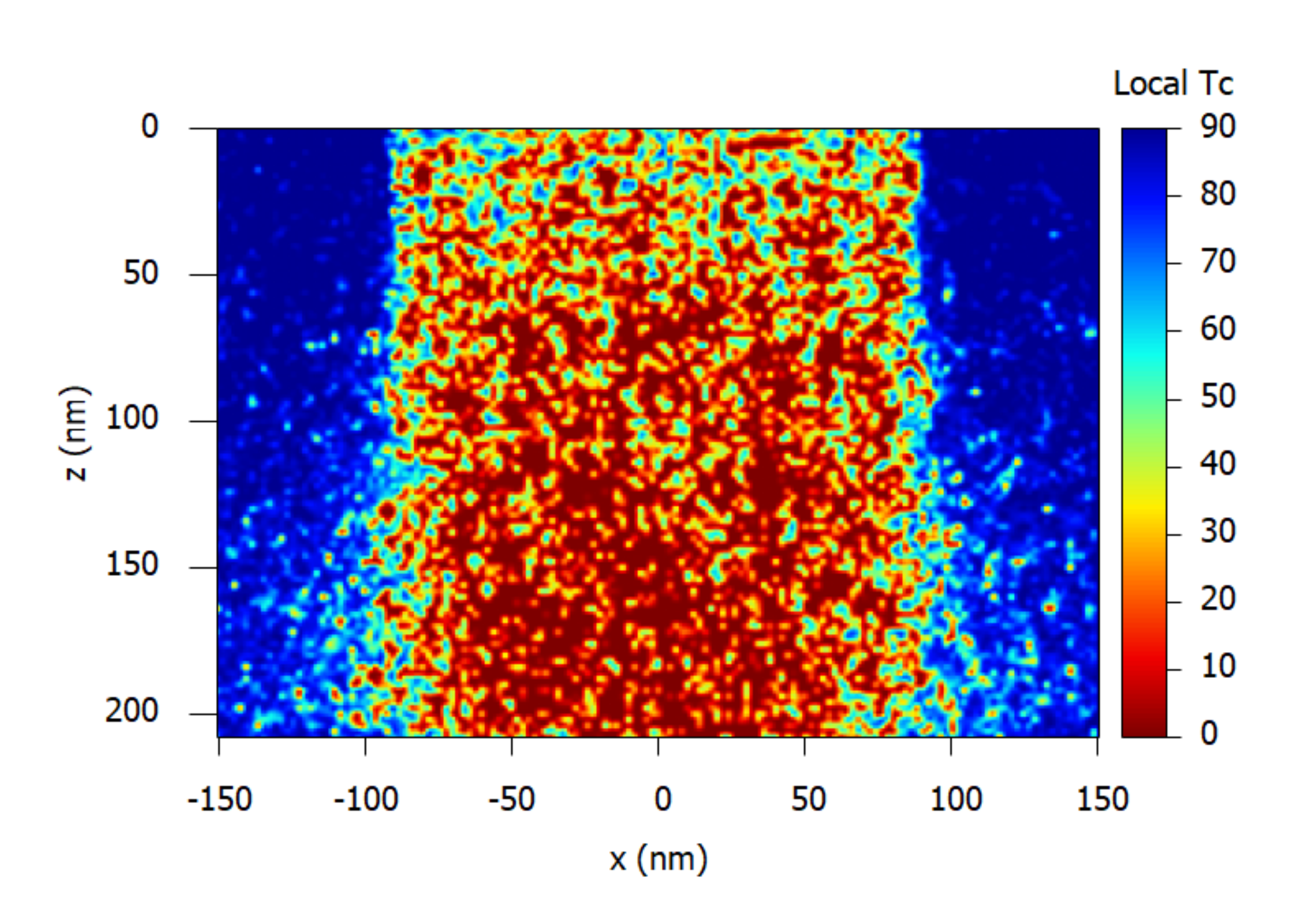}
        \subcaption{Cross-sectional view of the local $T_c$ profile.}
    \end{subfigure}
    \caption{Local $T_c$ profiles within and around a defect column produced by 75~keV He$^+$ ion irradiation of YBCO with a fluence of $3\times10^{15}\textrm{~cm}^{-2}$.}
    \label{fig:Tc3D}
\end{figure*}

As found in other studies with different irradiation parameters \cite{TOLP96,LESU90} the dependence of $T_\text{c-onset}$ with defect density roughly follows Eq.~(\ref{eq:pairbreaking}) with minor deviations \textbf{}near $T_\text{c-onset} \rightarrow T_0$ and $T_\text{c-onset} \rightarrow 0$. For this reason and the complexity of the implicit Eq.~(\ref{eq:pairbreaking}) we use a fit to the data for further interpolation, shown as a dashed line in Fig.~\ref{fig:dpa_Tc}, using $T_0=90$~K and the fixed constraint of going through the point (0,\ 90~K). With this function the local $T_c$ can be calculated from the 3D defect profiles in Fig.~\ref{fig:Tc}. The results are shown in Fig.~\ref{fig:Tc3D} as lateral $T_c$ profiles at various depths and as a cross-sectional view. When comparing the profiles of defects and local $T_c$ in the sample, it is remarkable that, even at the bottom of the sample, the region of suppressed superconductivity shows only marginal blurring and reproduces the shape of the incident ion irradiation well. Fig.~\ref{fig:Tc3D}(b) demonstrates that the channel of suppressed $T_c$ has rather sharp fringes and only marginally widens up with depth. Ideally, for experiments YBCO films with a thickness of only about 100~nm should be used or, alternatively the ion energy and consequently their fluence should be increased. These findings confirm the previous experimental results that indeed irradiation of YBCO with moderate-energy He$^+$ ions through a stencil mask allows for the fabrication of well defined 3D landscapes of the superconducting order parameter.

\section{Conclusions}

The 3D profile of the local critical temperature in a cuprate high-$T_c$ superconductor that was nano-patterned by ion irradiation has been examined. The interaction of 75~keV He$^+$ ions with a thin film of YBCO leads to displacements of mainly the oxygen atoms and creates point defects in the material. These defects cause a suppression of superconductivity and are deduced from simulations of the collision cascades with the program package SRIM/TRIM. Using previous experimental results for calibration we obtain a relation between the simulated defect density and the experimental $T_\text{c-onset}$. With these ingredients, the 3D pattern of the local $T_c$ in thin YBCO films can be visualized. It is found that masked ion irradiation through a stencil mask produces well-defined areas of suppressed $T_c$ with a lateral shape corresponding to the mask's holes with a resolution in the order of 10~nm. This structure persists up to 200~nm in depth with surprisingly low blurring. The present results support previous experimental findings of pronounced fluxon trapping effects that are caused by the particular $T_c$ landscape. Since ion irradiation through a stencil mask is a convenient fabrication method, our results support its potential usefulness for future quantum devices based on HTSCs.

\section*{Acknowledgments}

This work was supported by the European Cooperation in Science and Technology via COST Action CA16218 (NANOCOHYBRI).


\begin{thebibliography}{36}
\expandafter\ifx\csname natexlab\endcsname\relax\def\natexlab#1{#1}\fi
\providecommand{\url}[1]{\texttt{#1}}
\providecommand{\href}[2]{#2}
\providecommand{\path}[1]{#1}
\providecommand{\DOIprefix}{doi:}
\providecommand{\ArXivprefix}{arXiv:}
\providecommand{\URLprefix}{URL: }
\providecommand{\Pubmedprefix}{pmid:}
\providecommand{\doi}[1]{\href{http://dx.doi.org/#1}{\path{#1}}}
\providecommand{\Pubmed}[1]{\href{pmid:#1}{\path{#1}}}
\providecommand{\bibinfo}[2]{#2}
\ifx\xfnm\relax \def\xfnm[#1]{\unskip,\space#1}\fi
\bibitem[{Cardwell and Ginley(2003)}]{CARD03M}
\bibinfo{editor}{D.~A. Cardwell}, \bibinfo{editor}{D.~S. Ginley} (Eds.),
  \bibinfo{title}{Handbook of Superconducting Materials},
  \bibinfo{publisher}{IOP Publishing}, \bibinfo{address}{Bristol},
  \bibinfo{year}{2003}.
\bibitem[{Haugan et~al.(2004)Haugan, Barnes, Wheeler, Meisenkothen, and
  Sumption}]{HAUG04}
\bibinfo{author}{T.~Haugan}, \bibinfo{author}{P.~N. Barnes},
  \bibinfo{author}{R.~Wheeler}, \bibinfo{author}{F.~Meisenkothen},
  \bibinfo{author}{M.~Sumption}, \bibinfo{journal}{Nature}
  \bibinfo{volume}{430} (\bibinfo{year}{2004}) \bibinfo{pages}{867}.
  \DOIprefix\doi{10.1038/nature02792}.
\bibitem[{W\"ordenweber et~al.(2017)W\"ordenweber, Moshchalkov, Bending, and
  Tafuri}]{WORD17M}
\bibinfo{editor}{R.~W\"ordenweber}, \bibinfo{editor}{V.~Moshchalkov},
  \bibinfo{editor}{S.~Bending}, \bibinfo{editor}{F.~Tafuri} (Eds.),
  \bibinfo{title}{Superconductors at the Nanoscale, From Basic Research to
  Applications}, \bibinfo{publisher}{de Gruyter}, \bibinfo{address}{Berlin,
  Boston}, \bibinfo{year}{2017}. \DOIprefix\doi{10.1515/9783110456806}.
\bibitem[{Moshchalkov et~al.(2010)Moshchalkov, W\"ordenweber, and
  Lang}]{MOSH10M}
\bibinfo{editor}{V.~V. Moshchalkov}, \bibinfo{editor}{R.~W\"ordenweber},
  \bibinfo{editor}{W.~Lang} (Eds.), \bibinfo{title}{Nanoscience and Engineering
  in Superconductivity}, \bibinfo{publisher}{Springer},
  \bibinfo{address}{Heidelberg}, \bibinfo{year}{2010}.
  \DOIprefix\doi{10.1007/978-3-642-15137-8}.
\bibitem[{Tolpygo et~al.(1996)Tolpygo, Lin, Gurvitch, Hou, and
  Phillips}]{TOLP96}
\bibinfo{author}{S.~K. Tolpygo}, \bibinfo{author}{J.~Y. Lin},
  \bibinfo{author}{M.~Gurvitch}, \bibinfo{author}{S.~Y. Hou},
  \bibinfo{author}{J.~M. Phillips}, \bibinfo{journal}{Phys. Rev. B}
  \bibinfo{volume}{53} (\bibinfo{year}{1996}) \bibinfo{pages}{12462--74}.
  \DOIprefix\doi{10.1103/PhysRevB.53.12462}.
\bibitem[{Cui et~al.(1992)Cui, Xie, and Li}]{CUI92}
\bibinfo{author}{F.~Z. Cui}, \bibinfo{author}{J.~Xie}, \bibinfo{author}{H.~D.
  Li}, \bibinfo{journal}{Phys. Rev. B} \bibinfo{volume}{46}
  (\bibinfo{year}{1992}) \bibinfo{pages}{11182--5}.
  \DOIprefix\doi{10.1103/PhysRevB.46.11182}.
\bibitem[{Wang et~al.(1995)Wang, Hellebrand, B\"auerle, Strecker, Wortmann, and
  Lang}]{WANG95b}
\bibinfo{author}{X.~Z. Wang}, \bibinfo{author}{B.~Hellebrand},
  \bibinfo{author}{D.~B\"auerle}, \bibinfo{author}{M.~Strecker},
  \bibinfo{author}{G.~Wortmann}, \bibinfo{author}{W.~Lang},
  \bibinfo{journal}{Physica C} \bibinfo{volume}{242} (\bibinfo{year}{1995})
  \bibinfo{pages}{55--62}. \DOIprefix\doi{10.1016/0921-4534(94)02370-0}.
\bibitem[{Jorgensen et~al.(1990)Jorgensen, Pei, Lightfoot, Shi, Paulikas, and
  Veal}]{JORG90}
\bibinfo{author}{J.~D. Jorgensen}, \bibinfo{author}{S.~Pei},
  \bibinfo{author}{P.~Lightfoot}, \bibinfo{author}{H.~Shi},
  \bibinfo{author}{A.~P. Paulikas}, \bibinfo{author}{B.~W. Veal},
  \bibinfo{journal}{Physica C} \bibinfo{volume}{167} (\bibinfo{year}{1990})
  \bibinfo{pages}{571--8}. \DOIprefix\doi{10.1016/0921-4534(90)90676-6}.
\bibitem[{Stockinger et~al.(1998)Stockinger, Markowitsch, Lang, Kula, and
  Sobolewski}]{STOC98a}
\bibinfo{author}{C.~Stockinger}, \bibinfo{author}{W.~Markowitsch},
  \bibinfo{author}{W.~Lang}, \bibinfo{author}{W.~Kula},
  \bibinfo{author}{R.~Sobolewski}, \bibinfo{journal}{Eur. Phys. J. B}
  \bibinfo{volume}{2} (\bibinfo{year}{1998}) \bibinfo{pages}{301--11}.
  \DOIprefix\doi{10.1007/s100510050253}.
\bibitem[{Lang et~al.(2004)Lang, Enzenhofer, Peruzzi, Pedarnig, B\"auerle,
  Horner, Cekan, Platzgummer, and Loeschner}]{LANG04}
\bibinfo{author}{W.~Lang}, \bibinfo{author}{T.~Enzenhofer},
  \bibinfo{author}{M.~Peruzzi}, \bibinfo{author}{J.~D. Pedarnig},
  \bibinfo{author}{D.~B\"auerle}, \bibinfo{author}{C.~Horner},
  \bibinfo{author}{E.~Cekan}, \bibinfo{author}{E.~Platzgummer},
  \bibinfo{author}{H.~Loeschner}, \bibinfo{journal}{Inst. Phys. Conf. Ser.}
  \bibinfo{volume}{181} (\bibinfo{year}{2004}) \bibinfo{pages}{1549--1555}.
\bibitem[{{Cybart} et~al.(2014){Cybart}, {Yen}, {Cho}, {Huh}, {Glyantsev},
  {Yung}, {Moeckly}, {Beeman}, and {Dynes}}]{CYBA14a}
\bibinfo{author}{S.~A. {Cybart}}, \bibinfo{author}{P.~X.~T. {Yen}},
  \bibinfo{author}{E.~Y. {Cho}}, \bibinfo{author}{J.~U. {Huh}},
  \bibinfo{author}{V.~N. {Glyantsev}}, \bibinfo{author}{C.~S. {Yung}},
  \bibinfo{author}{B.~{Moeckly}}, \bibinfo{author}{J.~W. {Beeman}},
  \bibinfo{author}{R.~C. {Dynes}}, \bibinfo{journal}{IEEE Trans. Appl.
  Supercond.} \bibinfo{volume}{24} (\bibinfo{year}{2014})
  \bibinfo{pages}{1--5}. \DOIprefix\doi{10.1109/TASC.2014.2311400}.
\bibitem[{Haag et~al.(2014)Haag, Zechner, Lang, Dosmailov, Bodea, and
  Pedarnig}]{HAAG14}
\bibinfo{author}{L.~T. Haag}, \bibinfo{author}{G.~Zechner},
  \bibinfo{author}{W.~Lang}, \bibinfo{author}{M.~Dosmailov},
  \bibinfo{author}{M.~A. Bodea}, \bibinfo{author}{J.~D. Pedarnig},
  \bibinfo{journal}{Physica C} \bibinfo{volume}{503} (\bibinfo{year}{2014})
  \bibinfo{pages}{75--81}. \DOIprefix\doi{10.1016/j.physc.2014.03.032}.
\bibitem[{Vazquez-Mena et~al.(2015)Vazquez-Mena, Gross, Xie, Villanueva, and
  Brugger}]{VAZQ15}
\bibinfo{author}{O.~Vazquez-Mena}, \bibinfo{author}{L.~Gross},
  \bibinfo{author}{S.~Xie}, \bibinfo{author}{L.~Villanueva},
  \bibinfo{author}{J.~Brugger}, \bibinfo{journal}{Microelectron. Eng.}
  \bibinfo{volume}{132} (\bibinfo{year}{2015}) \bibinfo{pages}{236 -- 54}.
  \DOIprefix\doi{10.1016/j.mee.2014.08.003}.
\bibitem[{Villanueva et~al.(2011)Villanueva, Martin-Olmos, Vazquez-Mena,
  Montserrat, Langlet, Bausells, and Brugger}]{VILL11a}
\bibinfo{author}{L.~G. Villanueva}, \bibinfo{author}{C.~Martin-Olmos},
  \bibinfo{author}{O.~Vazquez-Mena}, \bibinfo{author}{J.~Montserrat},
  \bibinfo{author}{P.~Langlet}, \bibinfo{author}{J.~Bausells},
  \bibinfo{author}{J.~Brugger}, \bibinfo{journal}{IEEE Trans. Nanotechnol.}
  \bibinfo{volume}{10} (\bibinfo{year}{2011}) \bibinfo{pages}{940--6}.
  \DOIprefix\doi{10.1109/tnano.2010.2090171}.
\bibitem[{Kahlmann et~al.(1998)Kahlmann, Engelhardt, Schubert, Zander, Buchal,
  and Hollkott}]{KAHL98}
\bibinfo{author}{F.~Kahlmann}, \bibinfo{author}{A.~Engelhardt},
  \bibinfo{author}{J.~Schubert}, \bibinfo{author}{W.~Zander},
  \bibinfo{author}{C.~Buchal}, \bibinfo{author}{J.~Hollkott},
  \bibinfo{journal}{Appl. Phys. Lett.} \bibinfo{volume}{73}
  (\bibinfo{year}{1998}) \bibinfo{pages}{2354--6}.
  \DOIprefix\doi{10.1063/1.122459}.
\bibitem[{Katz et~al.(2000)Katz, Woods, and Dynes}]{KATZ00}
\bibinfo{author}{A.~S. Katz}, \bibinfo{author}{S.~I. Woods},
  \bibinfo{author}{R.~C. Dynes}, \bibinfo{journal}{J. Appl. Phys.}
  \bibinfo{volume}{87} (\bibinfo{year}{2000}) \bibinfo{pages}{2978--83}.
  \DOIprefix\doi{10.1063/1.372286}.
\bibitem[{Bergeal et~al.(2005)Bergeal, Grison, Lesueur, Faini, Aprili, and
  Contour}]{BERG05}
\bibinfo{author}{N.~Bergeal}, \bibinfo{author}{X.~Grison},
  \bibinfo{author}{J.~Lesueur}, \bibinfo{author}{G.~Faini},
  \bibinfo{author}{M.~Aprili}, \bibinfo{author}{J.~P. Contour},
  \bibinfo{journal}{Appl. Phys. Lett.} \bibinfo{volume}{87}
  (\bibinfo{year}{2005}) \bibinfo{pages}{102502}.
  \DOIprefix\doi{10.1063/1.2037206}.
\bibitem[{Swiecicki et~al.(2012)Swiecicki, Ulysse, Wolf, Bernard, Bergeal,
  Briatico, Faini, Lesueur, and Villegas}]{SWIE12}
\bibinfo{author}{I.~Swiecicki}, \bibinfo{author}{C.~Ulysse},
  \bibinfo{author}{T.~Wolf}, \bibinfo{author}{R.~Bernard},
  \bibinfo{author}{N.~Bergeal}, \bibinfo{author}{J.~Briatico},
  \bibinfo{author}{G.~Faini}, \bibinfo{author}{J.~Lesueur},
  \bibinfo{author}{J.~E. Villegas}, \bibinfo{journal}{Phys. Rev. B}
  \bibinfo{volume}{85} (\bibinfo{year}{2012}) \bibinfo{pages}{224502}.
  \DOIprefix\doi{10.1103/physrevb.85.224502}.
\bibitem[{Kang et~al.(2002)Kang, Burnell, Lloyd, Speaks, Peng, Jeynes, Webb,
  Yun, Moon, Oh, Tarte, Moore, and Blamire}]{KANG02a}
\bibinfo{author}{D.~J. Kang}, \bibinfo{author}{G.~Burnell},
  \bibinfo{author}{S.~J. Lloyd}, \bibinfo{author}{R.~S. Speaks},
  \bibinfo{author}{N.~H. Peng}, \bibinfo{author}{C.~Jeynes},
  \bibinfo{author}{R.~Webb}, \bibinfo{author}{J.~H. Yun},
  \bibinfo{author}{S.~H. Moon}, \bibinfo{author}{B.~Oh}, \bibinfo{author}{E.~J.
  Tarte}, \bibinfo{author}{D.~F. Moore}, \bibinfo{author}{M.~G. Blamire},
  \bibinfo{journal}{Appl. Phys. Lett.} \bibinfo{volume}{80}
  (\bibinfo{year}{2002}) \bibinfo{pages}{814--16}.
  \DOIprefix\doi{10.1063/1.1446998}.
\bibitem[{Blamire et~al.(2003)Blamire, Kang, Burnell, Peng, Webb, Jeynes, Yun,
  Moon, and Oh}]{BLAM03}
\bibinfo{author}{M.~G. Blamire}, \bibinfo{author}{D.~J. Kang},
  \bibinfo{author}{G.~Burnell}, \bibinfo{author}{N.~H. Peng},
  \bibinfo{author}{R.~Webb}, \bibinfo{author}{C.~Jeynes},
  \bibinfo{author}{J.~H. Yun}, \bibinfo{author}{S.~H. Moon},
  \bibinfo{author}{B.~Oh}, \bibinfo{journal}{Vacuum} \bibinfo{volume}{69}
  (\bibinfo{year}{2003}) \bibinfo{pages}{11--15}.
  \DOIprefix\doi{10.1016/s0042-207x(02)00303-2}.
\bibitem[{Lang et~al.(2006)Lang, Dineva, Marksteiner, Enzenhofer, Siraj,
  Peruzzi, Pedarnig, B\"auerle, Korntner, Cekan, Platzgummer, and
  Loeschner}]{LANG06a}
\bibinfo{author}{W.~Lang}, \bibinfo{author}{M.~Dineva},
  \bibinfo{author}{M.~Marksteiner}, \bibinfo{author}{T.~Enzenhofer},
  \bibinfo{author}{K.~Siraj}, \bibinfo{author}{M.~Peruzzi},
  \bibinfo{author}{J.~D. Pedarnig}, \bibinfo{author}{D.~B\"auerle},
  \bibinfo{author}{R.~Korntner}, \bibinfo{author}{E.~Cekan},
  \bibinfo{author}{E.~Platzgummer}, \bibinfo{author}{H.~Loeschner},
  \bibinfo{journal}{Microelectron. Eng.} \bibinfo{volume}{83}
  (\bibinfo{year}{2006}) \bibinfo{pages}{1495--1498}.
  \DOIprefix\doi{10.1016/j.mee.2006.01.091}.
\bibitem[{Zechner et~al.(2017)Zechner, Jausner, Haag, Lang, Dosmailov, Bodea,
  and Pedarnig}]{ZECH17a}
\bibinfo{author}{G.~Zechner}, \bibinfo{author}{F.~Jausner},
  \bibinfo{author}{L.~T. Haag}, \bibinfo{author}{W.~Lang},
  \bibinfo{author}{M.~Dosmailov}, \bibinfo{author}{M.~A. Bodea},
  \bibinfo{author}{J.~D. Pedarnig}, \bibinfo{journal}{Phys. Rev. Applied}
  \bibinfo{volume}{8} (\bibinfo{year}{2017}) \bibinfo{pages}{014021}.
  \DOIprefix\doi{10.1103/PhysRevApplied.8.014021}.
\bibitem[{Pedarnig et~al.(2010)Pedarnig, Siraj, Bodea, Puica, Lang, Kolarova,
  Bauer, Haselgr\"{u}bler, Hasenfuss, Beinik, and Teichert}]{PEDA10}
\bibinfo{author}{J.~D. Pedarnig}, \bibinfo{author}{K.~Siraj},
  \bibinfo{author}{M.~A. Bodea}, \bibinfo{author}{I.~Puica},
  \bibinfo{author}{W.~Lang}, \bibinfo{author}{R.~Kolarova},
  \bibinfo{author}{P.~Bauer}, \bibinfo{author}{K.~Haselgr\"{u}bler},
  \bibinfo{author}{C.~Hasenfuss}, \bibinfo{author}{I.~Beinik},
  \bibinfo{author}{C.~Teichert}, \bibinfo{journal}{Thin Solid Films}
  \bibinfo{volume}{518} (\bibinfo{year}{2010}) \bibinfo{pages}{7075--80}.
  \DOIprefix\doi{10.1016/j.tsf.2010.07.021}.
\bibitem[{Cybart et~al.(2015)Cybart, Cho, Wong, Wehlin, Ma, Huynh, and
  Dynes}]{CYBA15}
\bibinfo{author}{S.~A. Cybart}, \bibinfo{author}{E.~Y. Cho},
  \bibinfo{author}{T.~J. Wong}, \bibinfo{author}{B.~H. Wehlin},
  \bibinfo{author}{M.~K. Ma}, \bibinfo{author}{C.~Huynh},
  \bibinfo{author}{R.~C. Dynes}, \bibinfo{journal}{Nat. Nanotechnol.}
  \bibinfo{volume}{10} (\bibinfo{year}{2015}) \bibinfo{pages}{598}.
  \DOIprefix\doi{10.1038/nnano.2015.76}.
\bibitem[{Cho et~al.(2018{\natexlab{a}})Cho, Zhou, Cho, and Cybart}]{CHO18}
\bibinfo{author}{E.~Y. Cho}, \bibinfo{author}{Y.~W. Zhou},
  \bibinfo{author}{J.~Y. Cho}, \bibinfo{author}{S.~A. Cybart},
  \bibinfo{journal}{Appl. Phys. Lett.} \bibinfo{volume}{113}
  (\bibinfo{year}{2018}{\natexlab{a}}) \bibinfo{pages}{022604}.
  \DOIprefix\doi{10.1063/1.5042105}.
\bibitem[{Cho et~al.(2018{\natexlab{b}})Cho, Li, LeFebvre, Zhou, Dynes, and
  Cybart}]{CHO18a}
\bibinfo{author}{E.~Y. Cho}, \bibinfo{author}{H.~Li}, \bibinfo{author}{J.~C.
  LeFebvre}, \bibinfo{author}{Y.~W. Zhou}, \bibinfo{author}{R.~C. Dynes},
  \bibinfo{author}{S.~A. Cybart}, \bibinfo{journal}{Appl. Phys. Lett.}
  \bibinfo{volume}{113} (\bibinfo{year}{2018}{\natexlab{b}})
  \bibinfo{pages}{162602}. \DOIprefix\doi{10.1063/1.5048776}.
\bibitem[{M\"uller et~al.(2019)M\"uller, Karrer, Limberger, Becker,
  Schr\"oppel, Burkhardt, Kleiner, Goldobin, and Koelle}]{MULL19}
\bibinfo{author}{B.~M\"uller}, \bibinfo{author}{M.~Karrer},
  \bibinfo{author}{F.~Limberger}, \bibinfo{author}{M.~Becker},
  \bibinfo{author}{B.~Schr\"oppel}, \bibinfo{author}{C.~J. Burkhardt},
  \bibinfo{author}{R.~Kleiner}, \bibinfo{author}{E.~Goldobin},
  \bibinfo{author}{D.~Koelle}, \bibinfo{journal}{Phys. Rev. Applied}
  \bibinfo{volume}{11} (\bibinfo{year}{2019}) \bibinfo{pages}{044082}.
  \DOIprefix\doi{10.1103/PhysRevApplied.11.044082}.
\bibitem[{Zechner et~al.(2018)Zechner, Lang, Dosmailov, Bodea, and
  Pedarnig}]{ZECH18a}
\bibinfo{author}{G.~Zechner}, \bibinfo{author}{W.~Lang},
  \bibinfo{author}{M.~Dosmailov}, \bibinfo{author}{M.~A. Bodea},
  \bibinfo{author}{J.~D. Pedarnig}, \bibinfo{journal}{Phys. Rev. B}
  \bibinfo{volume}{98} (\bibinfo{year}{2018}) \bibinfo{pages}{104508}.
  \DOIprefix\doi{10.1103/PhysRevB.98.104508}.
\bibitem[{Ziegler et~al.(1985)Ziegler, Littmark, and Biersack}]{ZIEG85M}
\bibinfo{author}{J.~F. Ziegler}, \bibinfo{author}{U.~Littmark},
  \bibinfo{author}{J.~P. Biersack}, \bibinfo{title}{The stopping and range of
  ions in solids}, \bibinfo{publisher}{Pergamon, New York},
  \bibinfo{year}{1985}.
\bibitem[{Ziegler et~al.(2010)Ziegler, Ziegler, and Biersack}]{ZIEG10}
\bibinfo{author}{J.~F. Ziegler}, \bibinfo{author}{M.~D. Ziegler},
  \bibinfo{author}{J.~P. Biersack}, \bibinfo{journal}{Nucl. Instrum. Methods
  Phys. Res., Sect. B} \bibinfo{volume}{268} (\bibinfo{year}{2010})
  \bibinfo{pages}{1818--23}. \DOIprefix\doi{10.1016/j.nimb.2010.02.091}.
\bibitem[{Marwick and Clark(1989)}]{MARW89a}
\bibinfo{author}{A.~D. Marwick}, \bibinfo{author}{G.~J. Clark},
  \bibinfo{journal}{Nucl. Instrum. Methods Phys. Res., Sect. B}
  \bibinfo{volume}{37--38} (\bibinfo{year}{1989}) \bibinfo{pages}{910--6}.
  \DOIprefix\doi{10.1016/0168-583X(89)90326-1}.
\bibitem[{Legris et~al.(1993)Legris, Rullier-Albenque, Radeva, and
  Lejay}]{LEGR93}
\bibinfo{author}{A.~Legris}, \bibinfo{author}{F.~Rullier-Albenque},
  \bibinfo{author}{E.~Radeva}, \bibinfo{author}{P.~Lejay}, \bibinfo{journal}{J.
  Phys. I France} \bibinfo{volume}{3} (\bibinfo{year}{1993})
  \bibinfo{pages}{1605--15}. \DOIprefix\doi{10.1051/jp1:1993203}.
\bibitem[{Abrikosov and Gor'kov(1960)}]{ABRI60}
\bibinfo{author}{A.~A. Abrikosov}, \bibinfo{author}{L.~P. Gor'kov},
  \bibinfo{journal}{Zh. Eksp. Teor. Fiz.} \bibinfo{volume}{39}
  (\bibinfo{year}{1960}) \bibinfo{pages}{1781}. \bibinfo{note}{Sov. Phys. JETP
  {\bf 12}, 1243 (1961)}.
\bibitem[{Lesueur et~al.(1990)Lesueur, Nedellec, Bernas, Burger, and
  Dumoulin}]{LESU90}
\bibinfo{author}{J.~Lesueur}, \bibinfo{author}{P.~Nedellec},
  \bibinfo{author}{H.~Bernas}, \bibinfo{author}{J.~P. Burger},
  \bibinfo{author}{L.~Dumoulin}, \bibinfo{journal}{Physica C}
  \bibinfo{volume}{167} (\bibinfo{year}{1990}) \bibinfo{pages}{1--5}.
  \DOIprefix\doi{10.1016/0921-4534(90)90477-V}.
\bibitem[{Tolpygo et~al.(1996)Tolpygo, Lin, Gurvitch, Hou, and
  Phillips}]{TOLP96a}
\bibinfo{author}{S.~K. Tolpygo}, \bibinfo{author}{J.~Y. Lin},
  \bibinfo{author}{M.~Gurvitch}, \bibinfo{author}{S.~Y. Hou},
  \bibinfo{author}{J.~M. Phillips}, \bibinfo{journal}{Phys. Rev. B}
  \bibinfo{volume}{53} (\bibinfo{year}{1996}) \bibinfo{pages}{12454--61}.
  \DOIprefix\doi{10.1103/physrevb.53.12454}.
\bibitem[{N\'edellec et~al.(1989)N\'edellec, Lesueur, Traverse, Bernas,
  Dumoulin, and Laibowitz}]{NEDE89}
\bibinfo{author}{P.~N\'edellec}, \bibinfo{author}{J.~Lesueur},
  \bibinfo{author}{A.~Traverse}, \bibinfo{author}{H.~Bernas},
  \bibinfo{author}{L.~Dumoulin}, \bibinfo{author}{R.~B. Laibowitz},
  \bibinfo{journal}{J. Less Common Metals} \bibinfo{volume}{151}
  (\bibinfo{year}{1989}) \bibinfo{pages}{443--9}.
  \DOIprefix\doi{10.1016/0022-5088(89)90351-2}.

\end{thebibliography}
\end{document}